\documentstyle{article}
\begin{document}
\large
\title {\bf
Cabibbo favored hadronic decays of charmed baryons
in flavor SU(3)}
\author{
R. C. Verma and M. P. Khanna\\
\normalsize International Centre for Theoretical Physics, Trieste 34100.
{\bf Italy}\\
\normalsize and\\
\normalsize Centre for Advanced Study in Physics, Department of Physics, \\
\normalsize Panjab University, Chandigarh -160014 {\bf India}$^{*}$
}
\maketitle
\begin{abstract}
The two body Cabibbo favored hadronic decays of charmed
baryons $\Lambda^{+}_{c}$, $\Xi^{+}_{c}$ and $\Xi_{c}^{0}$ into an octet
or decuplet baryon and a pseudoscalar meson are examined in the SU(3)
symmetry scheme. The numerical estimates for decay widths and branching
ratios of some of the modes are obtained and are in good agreement with
experiment.
\end{abstract}
\vskip 0.5 cm
PACS number(s): 13.30.Eg, , 11.30.Hv, 11.40.Ha, 14.20.Kp \\
\vskip 1.0 cm
$^{*}$ Permanent Address
\newpage
\large
\section{Introduction}
The hyperon nonleptonic weak decays have so far evaded their complete
understanding. It is expected that the hadronic decays of the charmed and
heavier hadrons will be simpler and their study would help in the understanding
of the nonleptonic decay processes, in general. Upto now greater part of
theoretical effort to understand charm decays has been devoted to charmed
mesons.
Only recently, the study of hadronic two-body decays of the charmed baryons
has gained serious attention [1-7].
It is primarily due to the fact that some data on these decays has
started coming. The scarce data [8-10] which is available at present is already
beginning to discriminate between the theoretical models. It is our hope that
more and
more data will become available in the near future, and this data will provide
a new arena in which to study the standard model. Some very recent data [10]
prompts us to make a systematic analysis of the Cabibbo allowed decays of
charmed baryons in the framework of flavor SU(3) symmetry generated by $u,~d$
and
$s$ quarks.
\par
The two-body weak decay modes of hyperons have traditionally
been studied through the standard current algebra approach using the soft pion
theorem. It has been shown for quite some time, that though this approach
successfully reproducs the s- and p-wave amplitudes of the hyperons, and
their relative sign, it fails to predict their relative magnitudes [11,12]. To
have a better agreement between theory and experiment, the importance of
including the factorization contributions, which vanish in the soft-pion
limit, has been recognized [13].
\par
The weak decays of charmed baryons have been analyzed [1,7]
in the framework of soft-pion technique with the inclusion of factorization
terms. It had actually been expected that the factorization terms would
dominate. However, the observation of a few decays like
$\Lambda^{+}_{c} \rightarrow \Delta^{++}K^{-}/\Sigma^{0}\pi^{+}/\Xi^{0}K^{+}$
does not support this view and indicates the significance of pole diagrams
for charm changing decays. Further, the calculations of both the pole terms
and the factorization contribution have uncertainties associated with the
many parameters that have to be estimated.
\par
First, the soft-pion approach for the charm decays is suspect because the
meson emitted in the charm decays is far from being soft as there is a
lot of energy available in these decays. Second, the
baryon-baryon weak transitions involved in the commutator terms as well
as in the pole terms of the current algebra techniques have their own
uncertainties. The evaluation of the factorization involves the knowledge
of form factors which are not precisely known. The factorization contribution
turns out to be too large and has to be toned down arbitrarily to give
even a reasonable agreement with experiment.
\par These features have resulted in gross differences among the predictions
of various models and with experiment. The ratio of the experimental branching
ratio
for $ \Lambda \pi^{+}$ and $ p \bar K^{0}$ modes of $\Lambda_{c}^{+}$,
for example, is about 0.4 whereas the theoretical estimates [5] are
between 1 and 13. The $\Lambda^{+}_{c} \rightarrow \Sigma^{0}\pi^{+}/\Sigma^{+}
\pi^{0}$
modes do not get contribution from factorization term, while their  branching
ratios  are comparable to that of the $\Lambda \pi^{+}$ mode. Similarly,
the $\Lambda^{+}_{c} \rightarrow \Xi^{0} K^{+}$ mode does not get contribution
from
factorization, but the ratio of its branching fraction with that
of $\Lambda \pi^{+}$ mode is significant. These facts indicate that
weak decays of charm baryons are more complex than those of D-mesons, since
non-spectator processes are significant here.
Even by adjusting all the available parameters, the agreement of any
theoretical calculation done so far with the experimental observations
is far from satisfactory.
\par
The symmetry approach does have a number of parameters, but has the advantage
that it lumps the effects of all dynamical processes together. Since SU(3)
is a better symmetry than SU(4) for the charm hadrons [4], it is expected to
yield more reliable
results.  So we investigate the Cabibbo enhanced decays of charm =1 antitriplet
baryons in flavor SU(3) which we believe would be
valid in both the s- and p-wave modes.
\par For $B_{c} \rightarrow B + P $ mode, using $6^{*}$ domonance at the
SU(3) level, we have three parameters each for the s- and p-wave
amplitudes. First, we use the available data on $\Lambda^{+}_{c} \rightarrow
\Lambda \pi^{+} / \Sigma^{+} \pi^{0} /\Xi^{0}K^{+}$ decays to fix the
parameters and then make predictions on the remaining branching ratios and
asymmetries. Further, we attempt to relate p-wave charm baryon decays with
those of the hyperons following the approach of Altarelli, Cabibbo and
Maiani [14]. The predictions
obtained in the present analysis are consistent with the experimental values.
Particularly a small ratio of $Br (\Lambda^{+}_{c}\rightarrow \Lambda \pi^{+}
)/ Br(\Lambda^{+}_{c}\rightarrow p\bar K^{0})$ can be explained.
\par
We then extend our analysis to $B_{c}\rightarrow D + P$ decays. Here,
by employing the quark-line diagram approach of Kohara [3], we are able
to express amplitudes of all the $B_{c}$ decays in terms of only two
reduced amplitudes. Using $B(\Lambda^{+}_{c} \rightarrow \Delta^{++}K^{-}/
\Xi^{*0}K^{+})$ as input, branching ratio of the remaining decays are
predicted.
\section{ Weak Hamiltonian}
The general weak current $\otimes$ current weak Hamiltonian for Cabibbo
enhanced
 ($\Delta C= \Delta S = -1$) decays in terms of
the quark fields is
\begin{equation} H_{W}=
 {G_{F}\over {\sqrt 2}} V_{ud}V_{cs} (\bar u d)( \bar s c),
\label {(1)}\end{equation}
where $\bar q_{1}q_{2} \equiv \bar q_{1}\gamma_{\mu}(1-\gamma_{5})q_{2}$
represents the color-singlet combination.
If the QCD short distance effects are included the effective weak Hamitonian
becomes
\begin{equation} H_{W}=
 {G_{F}\over {\sqrt 2}} V_{ud}V_{cs}[ c_{1} (\bar u d) (\bar s c)
+ c_{2} (\bar s d) (\bar u c)],
\label {(1a)} \end{equation}
where the QCD coefficients $ c_{1} = {1 \over 2}(c_{+} + c_{-})$,
$c_{2} = {1 \over 2}(c_{+} - c_{-})$ and $c_{\pm}(\mu)= [\frac
{\alpha_{s}(\mu^{2})}
{\alpha_{s}(m_{W}^{2})}]^{d_{\pm}/2b}$ with $d_{-}= -2d_{+}=8$ and $b= 11-
{2\over 3}N_{f}$,
$N_{f}$ being the number of effective flavors, $\mu$ the mass scale. The value
of these QCD coefficients are difficult to assign, depending as they do on the
mass scale and $\Lambda_{QCD}$. For the charm sector, one usually takes,
$c_{1}= 1.2$, $c_{2} = -0.5$.
The effective Hamiltonian (2) then transforms as
an admixture of the $6^{*}$ and 15 representations of flavor SU(3).
\par
\section{$B_{c}({1 \over 2}^{+}) \rightarrow B({1\over 2}^{+}) + P(0^{-})$
decays}
We construct the effective Hamiltonian for the charm baryon decaying into
an octet baryon and a pseudoscalar meson by combining the final state baryon
octet
and meson octet multiplets into definite representations of SU(3), namely,
$1,~ 8_{S},~8_{A}~10,~10^{*},~ 27$. Then combining these representations with
charmed baryon antitriplet we construct $6^{*}$ and 15 Hamiltonian.
Equivalently,
we may construct SU(3) irreducible representations from the product of charm
baryon
antitriplet and weak Hamiltonian $6^{*}$ or 15 and contract it with the
irreducible
representations obtained in the product of final state baryon octet and meson
octet. The weak Hamiltonian is then given by
\begin{eqnarray}
H_{W}^{6^{*}} &=&{\sqrt 2} g_{8_{S}}\{ \bar B^{a}_{m}
P^{m}_{b}B^{n}H^{b}_{[n,a]}
     + \bar B^{m}_{b} P^{a}_{m}B^{n}H^{b}_{[n,a]}\}\
\nonumber \\
&{}&+{\sqrt 2}  g_{8_{A}}\{ \bar B^{a}_{m} P^{m}_{b}B^{n}H^{b}_{[n,a]}
           -\bar B^{m}_{b} P^{a}_{m}B^{n}H^{b}_{[n,a]} \}
\nonumber \\
&{}&+{{\sqrt 2}\over 2} g_{10^{*}}\{ \bar B^{a}_{b} P^{c}_{d}B^{b}H^{d}_{[a,c]}
                             + \bar B^{a}_{b} P^{c}_{d}B^{d}H^{b}_{[a,c]}
\nonumber \\
&{}&-{1\over 3} \bar B^{a}_{b} P^{c}_{a}B^{n}H^{b}_{[n,c]}
 + {1 \over 3} \bar B^{a}_{c} P^{c}_{d}B^{n}H^{d}_{[n,a]} \},
\label {(3)} \end{eqnarray}
\begin{eqnarray}
H_{W}^{15} &=& {{\sqrt 2}\over 2} h_{27} \{ \bar B^{a}_{b}
P^{c}_{d}B^{b}H^{d}_{(a,c)}
 + \bar B^{a}_{b} P^{c}_{d} B^{d} H^{b}_{(a,c)}
\nonumber \\
&{}&-{1\over 5} \bar B^{a}_{b} P^{c}_{a}B^{n}H^{b}_{(n,c)}
 - {1 \over 5} \bar B^{a}_{c} P^{c}_{d}B^{n}H^{d}_{(n,a)} \}
 \nonumber \\
&{}&+{{\sqrt 2}\over 2} h_{10} \{ \bar B^{a}_{b} P^{c}_{d}B^{b}H^{d}_{(a,c)}
 - \bar B^{a}_{b} P^{c}_{d}B^{d}H^{b}_{(a,c)}
 \nonumber \\
&{}& + {1\over 3} \bar B^{a}_{b} P^{c}_{a}B^{n}H^{b}_{(n,c)}
- {1 \over 3} \bar B^{a}_{c} P^{c}_{d}B^{n}H^{d}_{(n,a)} \}
\nonumber \\
&{}&+{\sqrt 2} h_{8_{S}}\{ \bar B^{a}_{m} P^{m}_{b}B^{n}H^{b}_{(n,a)}
     + \bar B^{m}_{b} P^{a}_{m}B^{n}H^{b}_{(n,a)}\}\
\nonumber \\
&{}&+{\sqrt 2} h_{8_{A}}\{ \bar B^{a}_{m} P^{m}_{b}B^{n}H^{b}_{(n,a)}
           -\bar B^{m}_{b} P^{a}_{m}B^{n}H^{b}_{(n,a)} \}.
\label {(4)} \end{eqnarray}
The QCD coefficients $c_{1}$ and $c_{2}$ are absorbed in the reduced amplitudes
$g$'s and $h$'s.
\subsection{The decay width and asymmetry formulas}
The matrix element for the baryon ${1 \over 2}^{+} \rightarrow {1\over 2}^{+} +
0^{-}$decay process is written as
\begin{equation}
  M = - \langle B_{f} P \vert H_{W}\vert B_{i} \rangle = i \bar u_{B_{f}}( A -
\gamma_{5}B) u_{B_{i}} \phi_{P} ,\label {(5)}
  \end{equation}
where A and B are parity violating (PV) and parity conserving (PC) amplitudes
respectively. The decay width is computed from
\begin{equation}
 \Gamma = C_{1} [ \vert A \vert^{2} + C_{2} \vert B \vert ^{2} ],
 \label {(6)} \end{equation}
where
$$ C_{1} = \frac { \vert {\bf q} \vert} {8 \pi} \frac { (m_{i} + m_{f})^{2} -
m_{P}^{2}} {m_{i}^{2}},$$
$$ C_{2} = \frac { (m_{i} - m_{f})^{2} - m_{P}^{2}}{(m_{i} + m_{f})^{2} +
m_{P}^{2}},$$
$$ \vert {\bf q} \vert = \frac {1}{2m_{i}} {\sqrt {[m_{i}^{2}- (m_{f} -
m_{P})^{2}][m_{i}^{2} - (m_{f} + m_{P})^{2}]}},$$
$m_{i}$ and $m_{f}$ are the masses of the initial and final baryons and $m_{P}$
is the mass of the emitted meson. Asymmetry parameter is given by
\begin{equation}
\alpha = \frac { 2 Re ( A \bar B^{*})}{ ( |A|^{2} + |\bar B |^{2})},
\label {(7)}\end{equation}
where $\bar B = \sqrt {C_{2}} B$.
\subsection{Decay amplitudes}
Choosing $H^{2}_{13}$ component of the weak Hamiltonian (\ref {(3)}) and (\ref
{(4)}),
decay amplitudes for various decays of antitriplet charmed baryons are obtained
and they are listed in Table 1. The C.G. coefficients occuring in this table,
are the same for PV as well as for PC modes. However, the reduced amplitudes
$g$'s and $h$'s will have different values for them. Thus in all, there exist
7 parameters in each PV and PC modes.
Perturbative corrections give rise to enhancement of coefficient of
$H^{6^{*}}_{W}$ over that of $H_{W}^{15}$. Consequently, it is possible, in
analogy with octet dominance in hyperon decays, that sextet dominance
may give reasonable results [2]. So
in order to reduce the number of parameters, we assume
6$^{*}$ dominance which  gives the following relations among the amplitudes:
\begin{equation}
\langle \Sigma^{0} \pi^{+} | \Lambda^{+}_{c} \rangle = -
\langle \Sigma^{+} \pi^{0} | \Lambda^{+}_{c} \rangle,    \label {(8)}
\end{equation}
\begin{equation}
\langle p \bar K^{0} | \Lambda^{+}_{c} \rangle = {{\sqrt 6}\over 2}\langle
\Lambda \pi^{+}| \Lambda^{+}_{c}\rangle
- {1 \over {\sqrt 2}} \langle \Sigma^{+} \pi^{0} | \Lambda^{+}_{c} \rangle,
\label {(9)}
\end{equation}
   \begin{equation}
\langle p \bar K^{0} | \Lambda^{+}_{c} \rangle =   \langle \Xi^{-} \pi^{+} |
\Xi^{o}_{c} \rangle,
 \label {(10)}\end{equation}
\begin{equation}
- \langle \Sigma^{0} \bar K^{0} | \Xi^{0}_{c} \rangle = {{\sqrt 3}\over
2}\langle \Lambda \pi^{+}| \Lambda^{+}_{c}\rangle
+ {1 \over  2} \langle \Sigma^{+} \pi^{0} | \Lambda^{+}_{c} \rangle,
\label {(11)}
\end{equation}
  \begin{equation}
- \langle \Lambda \bar K^{0} | \Xi^{0}_{c} \rangle = {1 \over 2}\langle \Lambda
\pi^{+}| \Lambda^{+}_{c}\rangle
- {{\sqrt 3}  \over  2} \langle \Sigma^{+} \pi^{0} | \Lambda^{+}_{c} \rangle.
\label {(12)} \end{equation}
Relation (\ref {(8)}) follows from the isospin subgroup of SU(3).
The above relations involve only two decay amplitudes    $\langle \Lambda
\pi^{+}| \Lambda^{+}_{c}\rangle$
 and  $ \langle \Sigma^{+} \pi^{0} | \Lambda^{+}_{c} \rangle$ on the right
hand side. We now add one more to obtain:
\begin{equation}
\langle \Sigma^{+} K^{-} | \Xi^{0}_{c} \rangle =   \langle \Xi^{0} K^{+} |
\Lambda^{+}_{c} \rangle,
\label {(13)} \end{equation}
 \begin{equation}
 \langle \Sigma^{+}  \eta_{8}  | \Lambda^{+}_{c} \rangle =
\sqrt {2 \over 3}  \langle \Xi^{0} K^{+} | \Lambda^{+}_{c} \rangle
- {1 \over {\sqrt 3}  } \langle \Sigma^{+} \pi^{0} | \Lambda^{+}_{c} \rangle,
\label {(14)}
\end{equation}
  \begin{equation}
- \langle \Xi^{0}  \pi^{0}  | \Xi^{0}_{c} \rangle =
 {1 \over {\sqrt 2} }  \langle \Xi^{0} K^{+} | \Lambda^{+}_{c} \rangle
-  \langle \Sigma^{+} \pi^{0} | \Lambda^{+}_{c} \rangle,
\label {(15)} \end{equation}
  \begin{equation}
 - \langle \Xi^{0}  \eta_{8}  | \Xi^{0}_{c} \rangle =
 {1 \over {\sqrt 6} }  \langle \Xi^{0} K^{+} | \Lambda^{+}_{c} \rangle
+ {1 \over {\sqrt 3}  } \langle \Sigma^{+} \pi^{0} | \Lambda^{+}_{c} \rangle,
 \label {(16)}
\end{equation}
 \begin{eqnarray}
&{}& - \langle \Xi^{0}  \pi^{+}  | \Xi^{+}_{c} \rangle =\langle \Sigma^{+}
\bar K^{0}  | \Xi^{+}_{c} \rangle=
\nonumber \\
&{}& -  \langle \Xi^{0} K^{+} | \Lambda^{+}_{c} \rangle
+ {{\sqrt 6} \over 2} \langle \Lambda \pi^{+}| \Lambda^{+}_{c}\rangle
+ {1 \over {\sqrt 2}  } \langle \Sigma^{+} \pi^{0} | \Lambda^{+}_{c} \rangle.
\label {(17)}
\end{eqnarray}
\subsection{Results and conclusion}
 Experimentally, branching ratios of all the Cabibbo enhanced
 $\Lambda^{+}_{c}\rightarrow B({1\over 2}^{+})+P(0^{-})$
decays (except $\Lambda^{+}_{c} \rightarrow \Sigma^{+}\eta^{\prime} )$) have
now been
measured [8-10]. Besides the decay asymmetries of
$\Lambda^{+}_{c} \rightarrow \Lambda \pi^{+}/\Sigma^{+} \pi^{0} $
have also become available. Following sets of PV and PC amplitudes (in units of
$G_{F} V_{ud}V_{cs}\times 10^{-2} GeV^{2})$ have been mentioned in a recent
CLEO
measurement [10],
$$ A(\Lambda^{+}_{c} \rightarrow \Lambda \pi^{+}) = -3.0^{+0.8}_{-1.2}  \qquad
\hbox{ or }   -4.3 ^{+0.8}_{-0.9},$$
$$ B(\Lambda^{+}_{c} \rightarrow \Lambda \pi^{+})= +12.7^{+2.7}_{-2.5}  \qquad
\hbox { or }  +8.9^{+3.4}_{-2.4}; $$
$$ A(\Lambda^{+}_{c} \rightarrow \Sigma^{+} \pi^{0})= +1.3^{+0.9}_{-1.1} \qquad
\hbox { or }  +5.4 ^{+0.9}_{-0.7},$$
\begin{equation}
 B(\Lambda^{+}_{c} \rightarrow \Sigma^{+} \pi^{0})= -17.3^{+2.3}_{-2.9}  \qquad
\hbox { or }   -4.1 ^{+3.4}_{-3.0}.
 \label {(18)} \end{equation}
 Relation (\ref {(8)}) immediately implies
\begin{equation}
 Br ( \Lambda^{+}_{c} \rightarrow \Sigma^{0} \pi^{+}) =
Br ( \Lambda^{+}_{c} \rightarrow \Sigma^{+} \pi^{0}).
\label{(19)}\end{equation}
Experimentally [8, 10] the L. H. S is $0.87 \pm 0.20 \%$ and R. H. S is
$0.87 \pm 0.22\%$. Also relation (\ref {(8)}) gives
 \begin{eqnarray}
 \alpha (\Lambda^{+}_{c} \rightarrow \Sigma^{0} \pi^{+}) &=&
\alpha (\Lambda^{+}_{c} \rightarrow \Sigma^{+} \pi^{0})
\nonumber \\
&=&- 0.45 \pm 0.31 \pm 0.06 .
\label {(20)}\end{eqnarray}
 However, relations (\ref {(9)}) to (\ref {(12)})
would lead to different values of the branching ratios and asymmetries
depending upon which set out of the four possibilities for
$ \Lambda^{+}_{c} \rightarrow \Lambda \pi^{+}$ and $ \Lambda^{+}_{c}
\rightarrow \Sigma^{+} \pi^{o}$
amplitudes is used as input. We carry out numerical analysis for all the four
choices and
find branching and asymmetries for various modes in the following ranges:
\begin{eqnarray}
Br (\Lambda_{c}^{+} \rightarrow p \bar K^{0}) &=& 2.74 \hbox{ to  } 3.51 \%,
\qquad
(\mbox {Expt.}~~2.1 \pm 0.4 \%)[8]
\nonumber \\
\alpha  (\Lambda_{c}^{+} \rightarrow p \bar K^{0}) &=& -0.72 \hbox {  to
}-0.99;
\label {(21)} \\
Br (\Xi_{c}^{0} \rightarrow \Lambda \bar K^{0}) &=& 0.69 \hbox { to } 0.87 \%,
\nonumber \\
\alpha (\Xi_{c}^{0} \rightarrow \Lambda \bar K^{0}) &=& -0.62 \hbox{ to } 0.86;
\label {(22)} \\
Br (\Xi_{c}^{0} \rightarrow \Xi^{-} \pi^{+}) &=& 1.30   \hbox{ to } 1.50 \%,
\nonumber \\
\alpha (\Xi_{c}^{0} \rightarrow \Xi^{-} \pi^{+}) &=& -0.75 \hbox { to }-0.99 ;
\label {(23)} \\
Br (\Xi_{c}^{0} \rightarrow \Sigma^{0} \bar K^{0}) &=& 0.06 \hbox { to }
0.19\%,
\nonumber \\
\alpha (\Xi_{c}^{0} \rightarrow \Sigma^{0} \bar K^{0}) &=& -0.67 \hbox { to  }
-0.87 \hbox{ or }+0.05 \hbox { to }+ 0.19.
\label {(24)}\end{eqnarray}
This is to be remarked that SU(3) symmetry predicts a large value of
branching ratio for $\Lambda^{+}_{c}\rightarrow p \bar K^{0}$ than
that of $\Lambda^{+}_{c}\rightarrow  \Lambda \pi^{+}$ in agreement with
experiment. The present data on $Br ( \Lambda^{+}_{c}\rightarrow  p\bar K^{0})$
seems to prefer the following choice
for the input:
\begin{eqnarray}
A(\Lambda^{+}_{c}\rightarrow \Sigma^{+}\pi^{0}) &=& + 5.4, \qquad
B(\Lambda^{+}_{c}\rightarrow \Sigma^{+}\pi^{0}) =-4.1;
\nonumber \\
A(\Lambda^{+}_{c}\rightarrow \Lambda \pi^{+}) &=&-3.0, \qquad
B(\Lambda^{+}_{c}\rightarrow \Lambda \pi^{+}) =+ 12.7 .
\label {(25)}\end{eqnarray}
To predict the remaining decays, we use the $\Lambda^{+}_{c}\rightarrow
\Xi^{0}K^{+}$.
Experimentally, only its branching ratio is known, $Br
(\Lambda^{+}_{c}\rightarrow \Xi^{0}K^{+})
\approx 0.34 \pm 0.09 \% ~[8]$. Ignoring the small kinematic difference, the
relation
(\ref {(13)}) gives
\begin{eqnarray}
Br (\Xi^{0}_{c}\rightarrow \Sigma^{+}K^{-})
&\approx &  Br (\Lambda^{+}_{c}\rightarrow \Xi^{0}K^{+})
=0.34 \pm 0.09 \%,
\nonumber \\
\alpha (\Xi^{0}_{c}\rightarrow \Sigma^{+}K^{-})
&\approx& \alpha  (\Lambda^{+}_{c}\rightarrow \Xi^{0}K^{+}).
\label{(26)}\end{eqnarray}
To be able to use other relations, we need the amplitude for the decay
$\Lambda^{+}_{c}\rightarrow \Xi^{0}K^{+}$ in both the PV and PC modes. The
measured branching ratio implies
\begin{equation}
|A(\Lambda^{+}_{c}\rightarrow \Xi^{0}K^{+})|^{2}
+ 0.055 |B(\Lambda^{+}_{c}\rightarrow \Xi^{0}K^{+})|^{2} = 14.42 .
\label {(27)}\end{equation}
The dynamical mechanisms considered for the charm baryon decay seem to
indicate that the PV mode of this decay is invariably strongly suppressed.
The decay can occur neither through the factorization nor from the equal time
commutator term of the current algebra framework. Even through the ${1\over
2}^{-}$
baryon pole, it acquires a negligibly small contributions [5]. Therefore, we
expect its asymmetry to be close to zero, which then gives
\begin{equation}
A(\Lambda^{+}_{c}\rightarrow \Xi^{0}K^{+}) \approx 0, \qquad
B(\Lambda^{+}_{c}\rightarrow \Xi^{0}K^{+}) = \pm 16.21.
\label {(28)}\end{equation}
Ignoring physical $\eta-\eta^{\prime} $ mixing, we then obtain
 \begin{eqnarray}
 Br (\Lambda^{+}_{c}\rightarrow \Sigma^{+}\eta) &=& 0.67 \%, \qquad
\alpha(\Lambda^{+}_{c}\rightarrow \Sigma^{+}\eta) = -0.95 \hbox{  for +ve
sign};
\nonumber \\
Br (\Lambda^{+}_{c}\rightarrow \Sigma^{+}\eta) &=& 0.45 \%, \qquad
\alpha(\Lambda^{+}_{c}\rightarrow \Sigma^{+}\eta) = +0.99 \hbox{  for -ve
sign}.
\label {(29)}\end{eqnarray}
where       $\Lambda^{+}_{c}\rightarrow \Lambda \pi^{+}$
and $\Lambda^{+}_{c}\rightarrow \Sigma^{+}\pi^{0}$ have been used from (\ref
{(25)}).
A recent CLEO measurement [9] gives
$${ Br(\Lambda^{+}_{c}\rightarrow \Sigma^{+}\eta)
\over  Br(\Lambda^{+}_{c} \rightarrow pK^{-}\pi^{+})} = 0.11\pm 0.03\pm 0.02
.$$
This measurement is consistent with both the theoretical predictions as
PDG data [8] gives $Br (\Lambda^{+}_{c}\rightarrow pK^{-}\pi^{+})= 4.4 \pm 0.6
\%$. So
we give branching ratio and decay asymmetry of the charm decays for both the
sets in Table 2. The values of the PC reduced amplitudes (in units of
$ G_{F} V_{ud} V_{cs} \times 10^{-2} GeV^{2}$) for these sets are:
\\
Set I: ~~  $B(\Lambda^{+}_{c}\rightarrow \Xi^{0}K^{+}) = - 16.21 $
 \begin{eqnarray}
 (g_{8_{S}})_{PC} &=& -1.09, ~~~(g_{8_{A}})_{PC} = -7.70,~~~(g_{10})_{PC}=
+28.81,
 \nonumber \\
(g_{8_{S}})_{PV} &=& +3.76, ~~~(g_{8_{A}})_{PV} = +3.80,~~~(g_{10})_{PV}=
+0.10;
 \label {(30)}\end{eqnarray}
Set II: ~~
{}~~  $B(\Lambda^{+}_{c}\rightarrow \Xi^{0}K^{+}) = + 16.21 $
 \begin{eqnarray}
 (g_{8_{S}})_{PC} &=& -17.30, ~~~(g_{8_{A}})_{PC} = -2.29,~~~(g_{10})_{PC}=
-3.61,
 \nonumber \\
(g_{8_{S}})_{PV} &=& +3.76, ~~~(g_{8_{A}})_{PV} = +3.80,~~~(g_{10})_{PV}=
+0.10.
 \label {(31)}\end{eqnarray}
Branching ratios of $\Xi^{+}_{c}$ decays show drastic difference between the
two sets.
Decay asymmetries of $\Xi_{c}^{0} \rightarrow \Xi^{0}+ \pi^{0}/\eta$,
though remaining large, have different  signs in the two cases.
\subsection{Relating charm baryon decays with hyperon decays}
The hyperon decays arise through
\begin{equation}
H_{W}= {G_{F}\over {\sqrt 2}} V_{ud}V_{us} [(\bar d u)(\bar u s)- (\bar d
c)(\bar c s)].
\label {(32)} \end{equation}
Under SU(3) symmetry, $H_{W}$ transforms like $ 8 \oplus 27$ representation and
the short distance effects enhance octet part over the 27 part, though the
enhancement factor falls short of experimental value. Using octet dominance
for hyperon decays, Altarelli, Cabibbo and Maiani [14]  related the charm
baryon decays with the
hyperon decays. They related the reduced amplitudes  $g$'s, and $h$'s
with those of the hyperon decays using CP-invariance at the SU(4) level.
In our phase convention, the relations are:
\begin{eqnarray}
(g_{8_{S}})_{PV} &=& {1\over 2} (g_{10})_{PV}= {1\over 2{\sqrt 6}}
[A(\Sigma_{+}^{+})+ {\sqrt 2} A(\Sigma^{+}_{0})],
\label {(33)}\\
(g_{8_{A}})_{PV} &=& {1\over 6{\sqrt 6}}[- A(\Sigma_{+}^{+})+ 5{\sqrt 2}
A(\Sigma^{+}_{0})],
\label {(34)}\\
(g_{10})_{PC} &=& {1\over {\sqrt 6}} B(\Sigma_{+}^{+})- B(\Lambda^{0}_{-}),
\label {(35)}\\
(g_{8_{A}})_{PC} &=& -{1\over 6{\sqrt 6}} B(\Sigma_{+}^{+})- {{\sqrt 3}\over 4}
B(\Sigma^{+}_{0})
+ {5 \over 12} B(\Lambda_{-}^{0})- {1\over 2}B(\Xi^{-}_{-}),
\label {(36)}\\
(g_{8_{S}})_{PC} &=& {1\over 2{\sqrt 6}} B(\Sigma_{+}^{+})- {{\sqrt 3}\over 4}
B(\Sigma^{+}_{0})
+ {3 \over 4} B(\Lambda_{-}^{0})- {1\over 2}B(\Xi^{-}_{-}).
\label {(37)}\end{eqnarray}
Unfortunately, the constraint (\ref {(33)}) forbids
$\Lambda_{c}^{+} \rightarrow \Lambda \pi^{+}$
 in PV mode and so would predict
$ \alpha (\Lambda_{c}^{+}\rightarrow \Lambda \pi^{+}) = 0$. Further, the
reduced
amplitudes obtained from these relations give very large branching ratio for
charm decays by a factor of 25 or so. In fact, $g_{8_{S}} = {1\over 2} g_{10}$
for PV mode is a typical consequence of SU(4) symmetry, giving
Iwasaki relation [15] for the hyperon decays,
 \begin{equation}
 \Lambda^{0}_{-}: \Sigma^{+}_{0}: \Xi^{-}_{-} = 1 : - {\sqrt 3} : 2,
 \label {(38)}\end{equation}
which is badly violated by experiment. The reason is that SU(4) symmetry
forbids factorization contributions in the PV mode, which is proportional
to the mass difference of the initial and final baryons. Therefore, such
relations among charm and uncharm sectors in PV mode are not reliable. However,
for PC mode, the relations (\ref {(35)}) to (\ref {(37)}) may still have some
meaning. Further, due to the
QCD modifications, the reduced amplitudes in charm sector are expected to be
lower than those needed for the hyperon decays [16]. Then,
lowering [16] the PC-reduced amplitudes $g_{8_{S}},~g_{8_{A}}$ and $g_{10}$
obtained
from (\ref {(35)}) to (\ref {(37)}) by a factor of 5, we use branching ratio of
$\Lambda^{+}_{c}\rightarrow    \Lambda \pi^{+}$ and
$\Lambda^{+}_{c}\rightarrow \Sigma^{+} \pi^{0}$ as input to determine the
PV-reduced amplitudes. The best set obtained
for branching ratios and decay asymmetry parameters for various charm-baryon
decays is given in Table 3. The values of the reduced amplitudes are:
 \begin{eqnarray}
 (g_{8_{S}})_{PC} &=& -17.72, ~~~(g_{8_{A}})_{PC} = -2.75,~~~(g_{10})_{PC}=
+3.81;
 \nonumber \\
(g_{8_{S}})_{PV} &=& +2.92, ~~~(g_{8_{A}})_{PV} = +3.19,~~~(g_{10})_{PV}=
+0.81.
\end{eqnarray}
These values and results obtained  match well with those given in col. (4) and
(5)
 of Table 3, and favor a
positive sign of $B(\Lambda^{+}_{c}\rightarrow \Xi^{0} K^{+})$.
\section{$B_{c}({1\over 2}^{+}) \rightarrow D({3\over 2}^{+}) + P(0^{-})$
decays}
In this section we examine the Cabibbo-favored decays of the anti-triplet
charm baryon ($B_{c}$) to a decuplet baryon (D) and a pseudoscalar meson (P).
The matrix element for the decay being defined as
\begin{equation}
\langle D, P| H_{W}|B_{c}\rangle = i q_{\mu} \bar {w}^{\mu}_{D}
(C - \gamma_{5} D)u_{B_{c}}\phi_{P},
\label {(39)}
\end{equation}
the decay rate and the asymmetry parameter for
$B({1 \over 2})^{+} \rightarrow D({3\over 2})^{+} + P(0)^{-}$ decay are given
by
 \begin{equation}
 \Gamma =  \frac { |{\bar{\bf q}}|^{3} m_{1}(m_{2} + E_{2})} { 6 \pi m_{2}^{2}}
 [ \vert C \vert^{2} +  \vert \bar D \vert ^{2} ], \label {(40)}
 \end{equation}
\begin{equation}
\alpha = \frac { 2 Re ( C \bar D^{*})}{ ( |C|^{2} + |\bar D |^{2})},
\label {(41)}
\end{equation}
where
$$
\bar D = \rho D, \qquad \rho = [(E_{2} - m_{2})/(E_{2} + m_{2})]^{1/2}.
$$
C and D are the p-wave (parity-conserving) and d-wave (parity-violating)
amplitudes respectively. $w_{\mu}$ is the Rarita-Scwhinger spinor representing
the spin $3/2^{+}$ baryon and $q_{\mu}$ is the four momentum of the emitted
meson.
\par
Following a procedure similar to that used for $B({1\over 2}^{+})\rightarrow
B({1\over 2}^{+}) + P(0^{-})$ decays, we construct the following
Hamiltonian for decuplet baryon emitting
decays:
\begin{equation}
H_{W}^{6^{*}}= {\sqrt 2} j_{8}(\epsilon_{mdb} \bar
D^{mnc}P^{d}_{n}B^{a}H^{b}_{[a,c]}),
\label {(42)} \end{equation}
\begin{eqnarray}
H_{W}^{15} &=& {\sqrt 2} k_{8}(\epsilon_{mpb} \bar
D^{mna}P^{p}_{n}B^{c}H^{b}_{(a,c)})
\nonumber \\
&{}& + {\sqrt 2} k_{10}(\epsilon_{mnd} \bar D^{mac}P^{n}_{b}B^{d}H^{b}_{(a,c)}
 - \epsilon_{mnb} \bar D^{mac}P^{n}_{d}B^{d}H^{b}_{(a,c)}
\nonumber \\
&{}&+ {2\over 3} \epsilon_{mnb} \bar D^{mdc}P^{n}_{d}B^{a}H^{b}_{(a,c)})
\nonumber \\
&{}& + {\sqrt 2} k_{27} (\epsilon_{mnd} \bar D^{mac}P^{n}_{b}B^{d}H^{b}_{(a,c)}
+ \epsilon_{mnb} \bar D^{mac}P^{n}_{d}B^{d}H^{b}_{(a,c)}\nonumber \\
&{}& - {2\over 5} \epsilon_{mnb} \bar D^{mdc}P^{n}_{d}B^{a}H^{b}_{(a,c)}),
\label {(43)}\end{eqnarray}
where $\epsilon_{abc}$ is the Levi-Civita symbol and $D_{abc}$ represents
totally symmetric decuplet baryons. Choosing $H^{2}_{13}$ component of
$H^{b}_{[a,c]}$ and $H^{b}_{(a,c)}$ tensors, we obtain the decay amplitudes
for various decay modes. These are shown in the
Table 4. In all there are 4 reduced amplitudes for each of the PV and PC modes.
Dynamically, these decays are relatively simpler than the ones considered
in the last section. It has been shown by Xu and Kamal [7] that factorization
terms do not contribute to these decays and that these decays arise only
through W-exchange diagrams. In fact, performing a quark diagram analyses,
Kohara [3] has observed that most of the quark diagrams, allowed for
$B({1\over 2}^{+}) \rightarrow B({1\over 2}^{+})+ P(0^{-})$ decays
are forbidden for $B({1\over 2}^{+}) \rightarrow D({3\over 2}^{+})+ P(0^{-})$
decays due to the symmetry property of the decuplet baryons. There exist
only two independent diagrams, which are expressed as:
\begin{equation}
A = d_{1}\bar D^{1ab}B_{[2,a]}M^{3}_{b}
+ d_{2} \bar D^{3ab} B_{[2,a]}M^{1}_{b},
\label {(44)} \end{equation}
where $B_{[a,b]}$ represents the $3^{*}$ baryon. In our notation, it amounts
to the folowing constraints:
\begin{equation}
k_{8} = {1\over 3} k_{10}, \qquad k_{27} = 0.
\label {(45)}\end{equation}
Following relations are obtained for PV as well as PC modes,
\begin{equation}
\langle \Xi^{*0} \pi^{+} |\Xi_{c}^{+}\rangle =
\langle \Sigma^{*+} \bar K^{0} |\Xi_{c}^{+}\rangle   =0,
\label {(46)}\end{equation}
\begin{equation}
\langle \Delta^{++} K^{-} |\Lambda_{c}^{+}\rangle =
{\sqrt 3}\langle \Delta^{+} \bar K^{0} |\Lambda_{c}^{+}\rangle =
{\sqrt 3}\langle \Sigma^{*+}  K^{-} |\Xi_{c}^{0}\rangle =
{\sqrt 6}\langle \Sigma^{*0}  \bar K^{0} |\Xi_{c}^{0}\rangle,
\label {(47)}\end{equation}
\begin{eqnarray}
&{}& {\sqrt 3} \langle \Xi^{*0} K^{+} |\Lambda_{c}^{+}\rangle =
{\sqrt 6}\langle \Sigma^{*+} \pi^{0} |\Lambda_{c}^{+}\rangle =
{\sqrt 6}\langle \Sigma^{*0}  \pi^{+} |\Lambda_{c}^{+}\rangle
\nonumber \\
&{}&=
{\sqrt 6}\langle \Xi^{*0}  \pi^{0} |\Xi_{c}^{0}\rangle
={\sqrt 3}\langle \Xi^{*-}  \pi^{+} |\Xi_{c}^{0}\rangle =
\langle \Omega^{-}  K^{+} |\Xi_{c}^{0}\rangle,
\label {(48)}\end{eqnarray}
\begin{equation}
 \langle \Sigma^{*+}  \eta_{8} |\Lambda_{c}^{+}\rangle =
\langle \Xi^{*0}  \eta_ {8} |\Xi_{c}^{0}\rangle=
{1 \over {\sqrt 6} }\langle \Xi^{*0} K^{+} |\Lambda_{c}^{+}\rangle
- {2 \over 3 {\sqrt 2}} \langle \Delta^{++} K^{-} |\Lambda_{c}^{+}\rangle.
\label {(49)}\end{equation}
Since W-exchange diagram contributions to the PV mode are generally small
and PV mode is suppressed due to the centrifugal barrier, we ignore them in
the present analysis. Experimentally [8], the following branching ratios are
known:
\begin{equation}
Br (\Lambda^{+}_{c}\rightarrow \Delta^{++}K^{-})= 0.7 \pm 0.4 \%,
\nonumber \end{equation}
\begin{equation}
Br (\Lambda^{+}_{c}\rightarrow \Xi^{*0}K^{+})= 0.23 \pm 0.09 \%,
\nonumber   \end{equation}
\begin{equation}
Br (\Lambda^{+}_{c}\rightarrow \Sigma^{*+}\eta )= 0.75 \pm 0.24\%.
\label{(50)}    \end{equation}
The last value has been taken from a recent CLEO measurements [9].
\begin{equation}
Br (\Lambda^{+}_{c}\rightarrow \Sigma^{*+}\eta )/
Br (\Lambda^{+}_{c}\rightarrow p K^{-}\pi^{+})
= 0.17 \pm 0.04 \pm 0.03, \nonumber
\end{equation}
with  $$Br (\Lambda^{+}_{c}\rightarrow pK^{-}\pi^{+})= 4.4 \pm 0.6 \%$$ from
PDG [8].
We employ $Br ( \Lambda^{+}_{c} \rightarrow \Delta^{++} K^{-} )$ and
$ Br ( \Lambda^{+}_{c} \rightarrow \Xi^{*0}K^{+})$ as input to fix,
$$ j_{8} = - 77.14, \qquad k_{8} = + 9.10~~ \hbox {(in units of }
G_{F}V_{ud}V_{cs}
\times 10^{-2} GeV^{2}),$$
which in turn give all other branching ratios. These are tabulated in Table 5.
For $\Lambda^{+}_{c}$, we expect $\Lambda^{+}_{c} \rightarrow \Sigma^{*} \pi $
modes
to be dominant and for $\Xi_{c}^{0}$ decay $\Xi^{0}_{c} \rightarrow
\Xi^{*0}\pi/
\Omega K$ modes are predicted to be dominant. Like in other theoretical models,
$\Xi_{c}^{+}$ decays in the present
analysis are also forbidden.
Their observations would indicate the presence of decay mechanism other than
the
W-exchange process.
Like $B({1\over 2}^{+}) \rightarrow B({1\over 2}^{+}) + P(0^{-})$ decays, here
also
one may like to relate these decays with decays of $\Omega^{-}$ hyperon. Since
$\Omega^{-}$-
decays do not involve W-exchange process, that comparison is not expected to
work.\\
\par
\noindent {\bf Acknowledgments}
\par
\noindent
The authors are grateful to Professor A. Salam, the President, and Professor
G. Virasoro, the Director, ICTP, Trieste,
UNESCO,  and Dr. S. Randjbar-Daemi   for hospitality at the ICTP, Trieste.
 
\newpage
\begin{table}
\begin{center}
\caption{$\Delta C= \Delta S = -1 $  decay amplitudes for charmed baryons.
}
\vskip 0.3 cm
\begin{tabular}{|c|c|c|} \hline
Decay  &  $H_{W}^{6^{*}}$     & $H_{W}^{15}$  \\
\hline \hline
 $ \Lambda_{c}^{+} \rightarrow p \bar K^{0}$ &  $- g_{8_{S}} -g_{8_{A}}+
{1\over 3}g_{10}$
 &  $h_{8_{S}} + h_{8_{A}}+ {1\over 3} h_{10}+ {2\over 5}h_{27}$\\
 $ \Lambda_{c}^{+}\rightarrow \Lambda \pi^{+}$ &  $ {1\over {\sqrt 6}}(-
2g_{8_{S}}+ g_{10})$
 &  ${1\over {\sqrt 6}} (2h_{8_{S}} - h_{10}- {6\over 5}h_{27})$\\
 $ \Lambda_{c}^{+}\rightarrow \Sigma^{+} \pi^{0} $ & ${1 \over {\sqrt
2}}(2g_{8_{A}}+ {1\over 3}g_{10})$
 &   ${1\over {\sqrt 2}} (- 2h_{8_{A}} + {1\over 3} h_{10}) $ \\
$ \Lambda_{c}^{+}\rightarrow \Sigma^{+} \eta_{8} $ &$ {1\over {\sqrt 6}}(-
2g_{8_{S}}- g_{10})$
 &  ${1\over {\sqrt 6}} (2h_{8_{S}} + h_{10}- {6\over 5}h_{27})$ \\
$ \Lambda_{c}^{+}\rightarrow \Sigma^{0} \pi^{+} $
&$ {1\over {\sqrt 2}}(- 2g_{8_{A}} -{1\over 3}  g_{10})$
 &  ${1\over {\sqrt 2}} (2h_{8_{A}} -{1\over 3} h_{10})  $\\
 $ \Lambda_{c}^{+} \rightarrow \Xi^{0} K^{+}$ & $ - g_{8_{S}}+g_{8_{A}}-
{1\over 3} g_{10}$
 &  $ h_{8_{S}} -h_{8_{A}}-{1\over 3} h_{10}+ {2\over 5}h_{27}$  \\
 {}&{}&{}\\
     &   &     \\
 $ \Xi_{c}^{+} \rightarrow \Xi^{0} \pi^{+}$ & $ - g_{10}$
 &  $ h_{27}  $ \\
 $ \Xi_{c}^{+} \rightarrow \Sigma^{+} \bar K^{0}$  & $  g_{10}$
 &  $ h_{27}  $ \\
       &   &     \\
 $ \Xi_{c}^{0} \rightarrow \Xi^{0} \pi^{0}$  & $ {1\over {\sqrt 2}} (
g_{8_{S}}+g_{8_{A}}+ {2\over 3} g_{10})$
 &  ${1\over {\sqrt 2}} (h_{8_{S}} + h_{8_{A}}+ {1\over 3}h_{10}- {3\over
5}h_{27}) $   \\
 $ \Xi_{c}^{0} \rightarrow \Xi^{0} \eta_{8}$ & ${1 \over {\sqrt 6}} (g_{8_{S}}
- 3g_{8_{A}})$
 & ${1 \over {\sqrt 6}} (h_{8_{S}} - 3h_{8_{A}}+ h_{10}- {3\over 5}h_{27})$
\\
 $ \Xi_{c}^{0} \rightarrow \Xi^{-} \pi^{+}$ & $- g_{8_{S}}-g_{8_{A}}+ {1\over
3} g_{10}$
 &  $ -h_{8_{S}} -h_{8_{A}}-{1\over 3} h_{10}- {2\over 5}h_{27} $    \\
 $ \Xi_{c}^{0} \rightarrow \Sigma^{+} K^{-}$ & $- g_{8_{S}}+g_{8_{A}}- {1\over
3} g_{10}$
 &  $- h_{8_{S}} +h_{8_{A}}+{1\over 3} h_{10} - {2\over 5}h_{27}$    \\
 $ \Xi_{c}^{0} \rightarrow \Sigma^{0} \bar K^{0}$ & $ {1 \over {\sqrt 2}}(
g_{8_{S}}- g_{8_{A}} - {2\over 3} g_{10})$
 &  ${1\over {\sqrt 2}} (h_{8_{S}} -h_{8_{A}}-{1\over 3} h_{10}- {3\over
5}h_{27} )$   \\
 $ \Xi_{c}^{0} \rightarrow \Lambda \bar K^{0}$ & ${1\over {\sqrt 6}}
(g_{8_{S}}+3 g_{8_{A}})$
 &  $ {1\over {\sqrt 6}} (h_{8_{S}} +3h_{8_{A}}- h_{10}- {3\over 5}h_{27}) $
\\
\hline
 \end{tabular}
 \end{center}
 \label {a}\end{table}

\newpage
\begin{table}
\begin{center}
\caption{Branching ratios and decay asymmetries of charmed baryons.}
\vskip 0.3 cm
\begin{tabular}{|c|c|c|c|c|} \hline
Decay  &  Br.  (\%)  & Asymmetry $\alpha$ &  Br.  (\%)  & Asymmetry $\alpha$
\\
  &  Set I  & Set I  &  SET II   & Set II \\
\hline \hline
 $ \Lambda_{c}^{+} \rightarrow p \bar K^{0}$ &  $2.74$ &  $-0.99$ &  $2.74$ &
$-0.99$ \\
 $ \Lambda_{c}^{+}\rightarrow \Lambda \pi^{+}$ &  $ 0.79^{\dag}$
 &  $-0.94^{\dag}$ &  $0.79^{\dag}$ &  $-0.94^{\dag}$ \\
 $ \Lambda_{c}^{+}\rightarrow \Sigma^{+} \pi^{0} $ & $0.87^{\dag}$
 &   $-0.45^{\dag} $ &  $0.87^{\dag}$ &  $-0.45^{\dag}$  \\
$ \Lambda_{c}^{+}\rightarrow \Sigma^{+} \eta $ & $ 0.45$
 &  $+0.99$ &  $0.67$ &  $-0.95$  \\
$ \Lambda_{c}^{+}\rightarrow \Sigma^{0} \pi^{+} $
&$ 0.87$
 &  $-0.45  $ &  $0.87$ &  $-0.45$  \\
 $ \Lambda_{c}^{+} \rightarrow \Xi^{0} K^{+}$ & $ 0.34^{\dag}$
 &  $ 0.00$  &  $0.34^{\dag}$ &  $-0.00$  \\
    &   &   &   &  \\
 $ \Xi_{c}^{+} \rightarrow \Xi^{0} \pi^{+}$ & $ 4.05 $
 &  $ +0.02  $ &  $0.06$ &  $-0.19$  \\
 $ \Xi_{c}^{+} \rightarrow \Sigma^{+} \bar K^{0}$  & $  4.23$
 &  $ +0.02 $ &  $0.07$ &  $-0.17$  \\
   &   &   &   &  \\
 $ \Xi_{c}^{0} \rightarrow \Xi^{0} \pi^{0}$  & $ 0.51$
 &  $ +0.71 $  &  $0.77$ &  $-0.99$    \\
 $ \Xi_{c}^{0} \rightarrow \Xi^{0} \eta$ & $0.20$
 & $-0.97$  &  $0.14$ &  $ +0.65$    \\
 $ \Xi_{c}^{0} \rightarrow \Xi^{-} \pi^{+}$ & $ 1.31$
 &  $ -0.96 $  &  $1.31 $ &  $ -0.96$    \\
 $ \Xi_{c}^{0} \rightarrow \Sigma^{+} K^{-}$ & $ 0.38$
 &  $ 0.00$  &  $0.38 $ &  $ 0.00$    \\
 $ \Xi_{c}^{0} \rightarrow \Sigma^{0} \bar K^{0}$ & $ 0.11$
 &  $ +0.05$  &  $0.11 $ &  $ +0.05$   \\
 $ \Xi_{c}^{0} \rightarrow \Lambda \bar K^{0}$ & $0.69$
 &  $ -0.86 $  &  $0.69 $ &  $ -0.86$    \\
\hline
 \end{tabular}
 \end{center}
 {\dag} input
 \label {b}\end{table}

\newpage
\begin{table}
\begin{center}
\caption{Branching ratios and asymmetries of charmed baryons using hyperon PC
amplitudes as input.}
\vskip 0.3 cm
\begin{tabular}{|c|c|c|} \hline
Decay  &  Br. ratio (\%)    & Asymmetry $\alpha$  \\
\hline \hline
 $ \Lambda_{c}^{+} \rightarrow p \bar K^{0}$ &  $2.70$     &  $-0.93$\\
 $ \Lambda_{c}^{+}\rightarrow \Lambda \pi^{+}$ &  $ 0.97^{\dag }$  &  $-0.66$\\
 $ \Lambda_{c}^{+}\rightarrow \Sigma^{+} \pi^{0} $ & $0.65^{\dag }$  &   $-0.38
$ \\
$ \Lambda_{c}^{+}\rightarrow \Sigma^{+} \eta $ & $ 0.48$ &  $-0.96$ \\
$ \Lambda_{c}^{+}\rightarrow \Sigma^{0} \pi^{+} $     &$ 0.65$  &  $-0.38  $\\
 $ \Lambda_{c}^{+} \rightarrow \Xi^{0} K^{+}$ & $ 0.24$   &  $ 0.00$  \\
       &   &     \\
 $ \Xi_{c}^{+} \rightarrow \Xi^{0} \pi^{+}$ & $ 0.11 $   &  $+0.94  $ \\
 $ \Xi_{c}^{+} \rightarrow \Sigma^{+} \bar K^{0}$  & $  0.10$  &  $+0.92 $ \\
       &   &     \\
 $ \Xi_{c}^{0} \rightarrow \Xi^{0} \pi^{0}$  & $ 0.55$  &  $-0.98 $   \\
 $ \Xi_{c}^{0} \rightarrow \Xi^{0} \eta $ & $0.11$   & $+0.67$    \\
 $ \Xi_{c}^{0} \rightarrow \Xi^{-} \pi^{+}$ & $ 1.15$   &  $-0.99 $    \\
 $ \Xi_{c}^{0} \rightarrow \Sigma^{+} K^{-}$ & $ 0.27$  &  $ 0.00$    \\
 $ \Xi_{c}^{0} \rightarrow \Sigma^{0} \bar K^{0}$ & $ 0.22$  &  $ +0.28$   \\
 $ \Xi_{c}^{0} \rightarrow \Lambda \bar K^{0}$ & $0.55$      &  $ -0.96 $    \\
\hline  $^{\dag}$ input
  \end{tabular}
 \end{center}

 \label {c}\end{table}

\newpage

\begin{table}
\begin{center}
\caption{$\Delta C=\Delta S=-1$ decay amplitudes of $B_{c}({1\over 2})^{+}
\rightarrow D({3\over 2})^{+}+ P(0^{-})$ decays.}
\vskip 0.3 cm
\begin{tabular}{|c|c|c|} \hline
Decay  &  $H_{W}^{6^{*}}$ &   $H_{W}^{15}$  \\
\hline \hline
 $ \Lambda_{c}^{+} \rightarrow  \Delta^{++} K^{-}$ &  $j_{1}  $ &
$ -k_{8} -{2\over 3}k_{10}  +{2\over 5}k_{27}$ \\
 $ \Lambda_{c}^{+}\rightarrow \Delta^{+} \bar K^{0}$ &  $ j_{1}/{\sqrt 3}$&
$ (-k_{8} -{2\over 3}k_{10}  +{2\over 5}k_{27})/{\sqrt 3}$  \\
 $ \Lambda_{c}^{+}\rightarrow \Sigma^{*+} \pi^{0} $ &
 $ - j_{1}/{\sqrt 6}$&
 $(k_{8} -{4\over 3}k_{10}  -{12\over 5}k_{27})/{\sqrt 6}$   \\
 $ \Lambda_{c}^{+}\rightarrow \Sigma^{*+} \eta_{8} $ &  $ - j_{1}/{\sqrt 2}$&
$ (k_{8} + {8\over 5}k_{27})/{\sqrt 2}$    \\
$ \Lambda_{c}^{+}\rightarrow \Sigma^{*0} \pi^{+} $ &
$ - j_{1}/{\sqrt 6}$&
$ (k_{8} -{4\over 3}k_{10}  -{12\over 5}k_{27})/{\sqrt 6}$  \\
$ \Lambda_{c}^{+}\rightarrow \Xi^{*0} K^{+} $     & $ - j_{1}/{\sqrt 3}$&
$ (k_{8} -{4\over 3}k_{10}  +{8\over 5}k_{27})/{\sqrt 3}$   \\
         &   &  \\
 $ \Xi_{c}^{+} \rightarrow \Sigma^{*+} \bar K^{0}$ & $ 0    $&$
(-4k_{27})/{\sqrt 3} $   \\
 $ \Xi_{c}^{+} \rightarrow \Xi^{*0} \pi^{+}$  & $0 $ & $(4 k_{27})/{\sqrt 3}$
\\
         &   &  \\
 $ \Xi_{c}^{0} \rightarrow \Sigma^{*+}  K^{-}$  &   $  j_{1}/{\sqrt 3}$&
$ (k_{8} -{4\over 3}k_{10}  +{8\over 5}k_{27})/{\sqrt 3} $  \\
   $ \Xi_{c}^{0} \rightarrow \Sigma^{*0} \bar K^{0}$ & $  j_{1}/{\sqrt 6}$&
$ (k_{8} -{4\over 3}k_{10}  -{12\over 5}k_{27})/{\sqrt 6}$   \\
 $ \Xi_{c}^{0} \rightarrow \Xi^{*0} \pi^{0}$ & $ - j_{1}/{\sqrt 6}$&
$ (- k_{8} -{2\over 3}k_{10}  -{18\over 5}k_{27})/{\sqrt 6} $  \\
 $ \Xi_{c}^{0} \rightarrow \Xi^{*0} \eta_{8} $ & $ - j_{1}/{\sqrt 2}$&
 $(- k_{8}+{2\over 3}k_{10}+ {2\over 5}k_{27})/{\sqrt 2} $  \\
 $ \Xi_{c}^{0} \rightarrow \Xi^{*-} \pi^{+}$ & $ - j_{1}/{\sqrt 3}$&
$ (- k_{8}-{2\over 3}k_{10}+ {2\over 5}k_{27})/{\sqrt 3} $  \\
 $ \Xi_{c}^{0} \rightarrow \Omega^{-}  K^{+}$ & $ - j_{1}$&
$- k_{8}-{2\over 3}k_{10}+ {2\over 5}k_{27}$
       \\
\hline
 \end{tabular}
 \end{center}
 \label {d}\end{table}

 \newpage
 \begin{table}
\begin{center}
\caption{Branching ratios of $B_{c}({1\over 2})^{+} \rightarrow D({3\over
2})^{+}+ P(0^{-})$ decays.}
\vskip 0.3 cm
\begin{tabular}{|c|c|}\hline
Decay  &  Branching ratio (\%)  \\
\hline \hline
 $ \Lambda_{c}^{+} \rightarrow  \Delta^{++} K^{-}$ &  $0.70^{\dag}$    \\
 $ \Lambda_{c}^{+}\rightarrow \Delta^{+} \bar K^{0}$ &  $ 0.23$  \\
 $ \Lambda_{c}^{+}\rightarrow \Sigma^{*+} \pi^{0} $ & $ 0.46$ \\
 $ \Lambda_{c}^{+}\rightarrow \Sigma^{*+} \eta $ & $0.30$   \\
$ \Lambda_{c}^{+}\rightarrow \Sigma^{*0} \pi^{+} $ & $ 0.46$ \\
$ \Lambda_{c}^{+}\rightarrow \Xi^{*0} K^{+} $     &$ 0.23^{\dag}$  \\
       &  \\
 $ \Xi_{c}^{+} \rightarrow \Sigma^{*+} \bar K^{0}$ & $ 0.0 $    \\
 $ \Xi_{c}^{+} \rightarrow \Xi^{*0} \pi^{+}$  & $  0.0$   \\
       &  \\
 $ \Xi_{c}^{0} \rightarrow \Sigma^{*+}  K^{-}$  & $ 0.13$    \\
 $ \Xi_{c}^{0} \rightarrow \Sigma^{*0} \bar K^{0}$ & $0.06$    \\
 $ \Xi_{c}^{0} \rightarrow \Xi^{*0} \pi^{0}$ & $ 0.26$  \\
 $ \Xi_{c}^{0} \rightarrow \Xi^{*0} \eta $ & $ 0.17$   \\
 $ \Xi_{c}^{0} \rightarrow \Xi^{*-} \pi^{+}$ & $ 0.51$   \\
 $ \Xi_{c}^{0} \rightarrow \Omega^{-}  K^{+}$ & $0.46$       \\
\hline
$^{\dag}$  input
 \end{tabular}
 \end{center}

 \label {e}\end{table}

\begin{thebibliography}{99}
\bibitem {} Short distance QCD enhancement
for the $6^{*}/15$ is roughly 2.6 times lower than that for 8/27.
\end{thebibliography}
\end{document}